\documentclass[prd,nofootinbib,preprint,floatfix]{revtex4}

\usepackage{graphicx}% Include figure files
\usepackage{dcolumn}% Align table columns on decimal point
\usepackage{bm}% bold math
\usepackage{comment}
\usepackage{caption}%new
\usepackage{subcaption}%new
\usepackage{float}%new

\usepackage[plainpages=false, colorlinks=true, anchorcolor=blue, linkcolor=blue, citecolor=blue, bookmarks=false]{hyperref}

\begin{document}

%\preprint{APS/123-QED}

\title{Can the  Efron-Petrosian Method  Recover the Inverse-Square Distance Law for  Simulated  Radio Pulsar Fluxes?}% Force line breaks with \\
\author{Sanjith \surname{A}}%
 \altaffiliation{sanjithajith2003a@gmail.com}
 \author{Shantanu \surname{Desai}}%
 \altaffiliation{shntn05@gmail.com}
\affiliation{Department of Physics, Indian Institute of Technology, Hyderabad, Kandi, Telangana-502284  India}

\date{\today}

\begin{abstract}
We test whether   the Efron-Petrosian (E-P) method can  recover the inverse-square law dependence of the radio pulsar flux, using a   synthetic catalog generated according to the specifications of the Parkes Multi-beam survey using the {\tt PsrPopPy} software.  We find that the E-P  method cannot automatically reproduce  the inverse-square  scaling law  for the radio pulsar flux, even though this scaling is built into the synthetic data by construction,  except over a narrow range of flux thresholds, and even here we don't get pristine agreement. The main reason for the deviation is that the synthetic radio pulsar  catalog is truncated based on a cut on the pulsar signal to noise ratio (SNR), which has a non-linear dependence on the flux along with considerable scatter. We show that the disagreement  is exacerbated as we raise the SNR threshold.
 We then demonstrate that if we create a synthetic catalog based on a flux cut (instead of an SNR-based threshold), we can recover the true distance exponent, with an accuracy ranging from pristine agreement to within $\pm 1 \sigma$ depending on the chosen flux threshold. Therefore, the E-P method cannot be used to determine the correct scaling of radio pulsar flux with distance. Our results also demonstrate that  the E-P method breaks down when the detection threshold scales non-linearly with flux along with scatter.

\end{abstract}
\maketitle

\section{Introduction}

Pulsars are rapidly spinning neutron stars, that produce pulsed radio emissions with rotation speeds  spanning  from a few   milliseconds up to several seconds   and    magnetic fields ranging from $10^8$ to $10^{14}$ G~\cite{handbook,Reddy}.  Pulsars serve as exceptional natural laboratories for a wide range of studies in physics and astrophysics (from solar system to cosmology)~\cite{Blandford92,KaspiKramer}. Some examples include  stellar evolution~\cite{Lorimer08}, dark matter contribution along the  line of sight~\cite{Kahya}, tests of equivalence principle~\cite{DesaiKahyaCrab}, structure of interstellar medium~\cite{He25}, coronal mass ejections from the Sun~\cite{Shaswat},  tests of  Lorentz Invariance violation~\cite{DesaiLIV}, etc.

In recent years, Ardavan has carried out multiple analyses   applying the  Efron-Petrosian (E-P) method~\cite{EP,Petrosian02} 
on  gamma-ray fluxes  of pulsars~\cite{Ardavanfermi,Ardavan23} and   X-ray emission of magnetars~\cite{Ardavanmagnetar,Ardavan24}.  These results showed  that  the scaling of the X-ray and    gamma-ray    flux density  ($F$) with the distance ($D$) according to  $F \propto D^{-3/2}$,  is favored at higher levels of significance compared to the scaling of the inverse-square law of $F \propto D^{-2}$.  This in turn provides support to the high energy pulsar emission mechanism  proposed in ~\cite{Ardavan21}.
Since the full details of the pulsar emission mechanism at radio wavelength are not completely understood~\cite{Melrose}, it would be worthwhile to apply the same techniques to radio pulsar fluxes, to probe the dependence of radio pulsar flux with distance.

Therefore, motivated by these considerations,   we applied the  E-P technique
to radio pulsar flux data  at 1400 MHz from the  Parkes Multi-beam survey data~\cite{PMBS}, in order to look for violations of inverse-square law scaling of the pulsar flux~\cite{Pragna} (M23, hereafter).
This E-P based analysis of the the  radio pulsar flux data  could not discern a clear correlation between the pulsar flux and distance for any index, unlike the analyses in Ardavan's works. Previously, the stepwise maximum likelihood method~\cite{SWML} has also been applied to pulsar radio fluxes to deduce the scaling of the pulsar flux with distance. Although a violation of the inverse-square   law for the radio pulsar flux was claimed~\cite{Singleton} based on this method, this result could not be independently reproduced~\cite{Desai16}.

However,  M23 did not demonstrate   that one could recover 
the correct inverse-square law scaling of the pulsar flux with distance ($F \propto 1/D^2$) using a synthetic radio  pulsar population, where the telescope and survey-related effects were included.  Demonstrating this for a control sample would be the  necessary first step before applying it  to the real data.

%Previously,  there have been claims that the radio pulsar fluxes do not follow an inverse square law in models involving superluminal polarization currents~\cite{Ardavan09}. Such a model predicts an inverse linear scaling of the flux with distance. A claim was also made in literature  that the  radio pulsar fluxes {\it do}  scale inversely with the  first power of  distance according to ($F \propto D^{-1}$)~\cite{Singleton,Middleditch} based on the application of the Stepwise Maximum Likelihood Method~\cite{SWML}, in accord with the theoretical model proposed in ~\cite{Ardavan09}. However, this result could not be confirmed  using an independent analysis~\cite{Desai16}. 

Therefore, in this work, we apply the E-P method to a synthetic pulsar population generated using the {\tt psrPoppy}~\cite{PsrPopPy,Bagchi20} package, to confirm whether the E-P correctly gives us the correct inverse-square law scaling.  The outline of this manuscript is as follows. We recap the E-P method and its myriad applications in the literature in Sect.~\ref{sec:EP}. 
The generation of the  simulated  radio pulsar catalog is described in Sect.~\ref{sec:simulations}.
The results from our analysis  can be found in Sect.~\ref{sec:analysis}. Corresponding results using a flux-limited catalog can be found  in Sect.~\ref{sec:fluxcut}.
We conclude in Sect.~\ref{sec:conclusions}. All logarithms in this manuscript are to the base 10.

\section{Primer on Efron-Petrosian method}
\label{sec:EP}
The E-P method has been widely used  throughout astrophysics  and cosmology  to account for selection biases or evolution  in flux (or magnitude)-limited or truncated samples~\cite{EP,Petrosian02}. 
This method has been used  to study a wide variety of astronomical phenomena from extra-galactic astronomy (eg. quasars) to solar system (eg. asteroids) for a diverse range of applications. A non-exhaustive list of such applications can be found in  ~\cite{LeePetrosian, Maloney,Wheatland,Kocevski,Dainotti13,Dainotti15,DPB21,Dainotti22,Bargiacchi23,Champati25,Guo25,Khatiya25,ChampatiPetro}. In most of these works, the E-P method has been used to determine the scaling between the two   observables for  flux-limited datasets. In this work, we turn the question around to investigate whether the  E-P method can be used to deduce  the correct  scaling ($\propto 1/D^2$) with the distance of the pulsar radio flux. We also  use the same notation as that used in  Ardavan~\cite{Ardavanfermi}.
%This method has been used for a variety of science goals such as a probe of luminosity evolution, checking for intrinsic correlations between two astrophysical variables, precision tests of cosmological models, search for  cosmological time dilation, tests of standard candles, etc.  We provide a brief description of  the EP method. More details about this technique can be found in the aforementioned references.

Consider a flux-limited catalog containing flux measurements ($F$) obtained from a series of observations. Assuming  that the pulsar flux scales with the distance ($D$) according to $F \propto D^{-\alpha}$, the isotropic luminosity ($L$) of the pulsar is given in terms of $F$ and $D$ according to~\cite{Ardavan23}:
\begin{equation}
L = 4 \pi l^2 \left(\frac{D}{l}\right)^{\alpha} F .
\label{eq:L}
\end{equation}
As discussed in M23 or ~\cite{Ardavan21}, $l$ is a constant with the dimensions of distance, primarily serving as the normalization factor.  For the inverse-square law, this leads to  the familiar relation  $L=4 \pi D^2 F$. We note however that in pulsar literature, instead of the conventional luminosity, one usually defines a pseudo-luminosity which omits the $4\pi$ factor in Eq.~\ref{eq:L}~\cite{Bagchi}, since the beam geometry is not known.  Therefore, similar to M23, we use the pseudo-luminosity for our E-P analysis and omit the $4\pi$ factor from now on.   We also note that throughout the remainder of the manuscript, the term luminosity refers to the pseudo-luminosity.

If we now assume that a  pulsar survey is limited by a flux   threshold $S_{th}$, then the characteristic luminosity cutoff $L_{th}$ scales with $D$ as:
\begin{equation}
\log L_{th} = \log[  l^{2-\alpha} S_{th}] + \alpha \log D
\end{equation}
For each distance-luminosity pair  ($\log D_i$, $\log L_i$), one can determine  a corresponding  set  of luminosity-distance ($L,D$) points in rest of the dataset  defined by: 
\begin{eqnarray}
  \log D &\leq& \log D_i~\rm{for}~i=1...n.  \\ \rm{and} 
  \log L &\geq& \log [  l ^{2-\alpha} S_{th}] +   \alpha \log D ,
\end{eqnarray}
where $n$ denotes the number of pulsars not excluded by the flux threshold. All ($D,L$) pairs that fulfill the  above criteria are often referred to  as  ``associated''~\cite{Dainotti22} or ``comparable''~\cite{Ardavanfermi}  set relative to ($D_i,L_i$). An illustrative plot showing all the values of $D$ associated with a given  pair of $D_i$  and $L_i$ can be found in Fig. 3 of M23 or Fig. 3 in ~\cite{Ardavan23}. The total number of  associated pairs corresponding to ($D_i,L_i$) is denoted by $N_i$. Next, we determine the rank ($\mathcal{R}_i$)  of this point  using $L_i$,  relative to  its associated set of points, when  sorted in  ascending order.

The E-P technique then calculates the following normalized statistic,  which is related to the  Kendall-$\tau$ statistic-for all data points ($n$) exceeding the flux threshold:
\begin{equation}
\tau = {{\sum \limits_{i=1}^n{(\mathcal{R}_i-\mathcal{E}_i)}} \over {\sqrt{\sum \limits_{i=1}^n{\mathcal{V}_i}}}}, 
\end{equation}
where $\mathcal{E}_i=\frac{1}{2}(N_i+1)$ and  $\mathcal{V}_i=\frac{1}{12}(N_i^2-1)$. 
The hypothesis that  $L$ and $D$ are independent of each other is therefore contingent on the absolute value of $\tau$.
If $L$ and $D$ are independent, then the value of $\tau$ is expected to be close to  zero. In contrast,  if they are correlated, $\tau$ has significantly higher values, allowing the hypothesis of independence to be rejected with high statistical significance.
This hypothesis of independence between distance and luminosity can be quantitatively assessed by computing a $p$-value, as described in~\cite{EP}:
\begin{equation}
p = \left(\frac{2}{\pi}\right)^{1/2} \int_{|\tau|}^{\infty} \exp(-x^2/2) dx
\label{eq:pvalue}
\end{equation}
In terms of  significance, one can reject the hypothesis that $L$ and $D$ are independent at $Z$-score values  equal to $n\sigma$  if $|\tau| > n$. In the literature, the E-P method has been applied in multiple ways. One approach involves scaling  the luminosity with a power-law of distance or redshift, and determining  the distance  exponent  for which $\tau=0$ (or $|\tau|<1$)~\cite{Dainotti13}, corresponding to the corrected luminosity being independent of distance~\cite{Bargiacchi23}. Alternately, one can evaluate  the relative significance of independence for different distance exponents. This is the methodology used in the recent works by Ardavan and also M23. We also note that the original EP paper~\cite{EP} does not provide any specific recommendations about the choice of  flux threshold to use for computing $\tau$. The early applications of E-P method used  the instrumental limiting sensitivity for the flux threshold~\cite{Maloney}. However, because this value  can sometimes be  too low,  a higher  flux threshold is often adopted~\cite{Dainotti13}. In recent works by Ardavan and P23, E-P statistics were evaluated at multiple flux thresholds, where each threshold was chosen such that the truncated and the full data set are drawn from the same distribution~\cite{Ardavan24}. This was tested using the Kolmogorov-Smirnov (KS)  test~\cite{Ardavan24,DPB21}.

It has also been recently pointed out  that one only gets physically meaningful results for the E-P method when the detection thresholds are chosen near the peak of the dataset histogram~\cite{Bryant}.  Numerous examples from literature have also been provided in ~\cite{Bryant} demonstrating that incorrect selection of flux thresholds can lead to misleading results.  In this work, we test the E-P method for four different thresholds within a given dataset following ~\cite{Ardavanfermi} and M23.

\section{Generation of synthetic pulsar catalog}
\label{sec:simulations}
For this analysis, we generate the synthetic pulsar population using the {\tt PsrPoppy} simulation software. This software generates a synthetic pulsar population  and builds up over previous such efforts~\cite{Bhatta92,Kaspi,Ridley,Lorimer06}. We run {\tt PsrPopPy} in snapshot mode, so that it outputs the current pulsar population.  The pulsar catalogs generated from this package  have been shown to agree with the results of different surveys. 
This software  has multiple options for the Galactic radial distribution of pulsars ranging from isotropic distribution
to distribution along the Galactic plane to more complicated distributions~\cite{Lorimer06,Yusifov}. For our analysis,  the pulsar distances were generated using the  galactic population model in ~\cite{Lorimer06} (using the  {\tt lfl06} command-line option), which uses a two-sided exponential with a scale height of 330 pc. We used the NE2001 electron density model~\cite{NE2001}.
We have used the default options  for the other pulsar parameters such as luminosity, spectral index, initial period, pulsar spin down, beam alignment, braking index, maximum pulsar age, magnetic field model, scattering distribution, which can be found in Table 2 of ~\cite{PsrPopPy}. Pulsar radio fluxes  are related to the 
(pseudo-)luminosity using the inverse-square law ($F=L/d^2$). Therefore, in order for the E-P technique to recover the correct inverse-square law scaling~\footnote{From now on, the inverse-square law discussed in the manuscript refers to the relation $F=L/d^2$.}, one should obtain $\tau=0$ for  the exponent $\alpha=2$.

For this analysis, we use the {\tt python3} version of {\tt PsrPopPy}\footnote{\url{https://github.com/mohak300501/PsrPopPy}}. This includes some of the bug fixes to the original {\tt psrPopPy} pointed out in ~\cite{Bagchi20}.
We generated the synthetic catalog of  2000 pulsars corresponding to the specifications of the  Parkes Multi-beam survey~\cite{PMBS}. For posterity, we provide the full command line options used to generate  the Parkes Multi-beam survey based population.
%\begin{minted}[breaklines, fontsize=\small]{bash}
\begin{verbatim}
 populate.py -n 2000 -surveys PMSURV -dm ne2001 -rdist lfl06 -o pop.model   
 dosurvey.py -f pop.model -surveys PMSURV
\end{verbatim}
The first command generates the population model corresponding to specifications of the Parkes Multi-beam survey with a lognormal distribution of luminosity. The second command runs the generated population through the Parkes Multi-beam survey.
We note that both these commands check against the default signal to noise ratio (SNR) threshold, before saving the output pulsars and continues until the desired number of pulsars (2000) are saved. For Parkes Multi-beam survey this SNR threshold is equal to 9.0.
Note that we have not included the effects caused by scintillation which have recently been included in {\tt PsrPoppy}. The final output of {\tt PsrPoppy} consists of pulsar period, dispersion measure, pulse width, spectral index, true distance, $X$, $Y$, $Z$ coordinates, and finally the flux and luminosity at 1400 MHz. Among these,  we only need the flux and the true distance for our analysis. In the Appendix, we shall also test the E-P method with a power-law distribution for the luminosity as well as with a synthetic population 
with flux scaling laws that differ from the inverse-square law.

\section{Analysis}
\label{sec:analysis}
\subsection{Results from Applications of E-P analysis on synthetic population of pulsars.}
\begin{figure}[hbt!]
    \centering
    \includegraphics[width=0.5\linewidth]{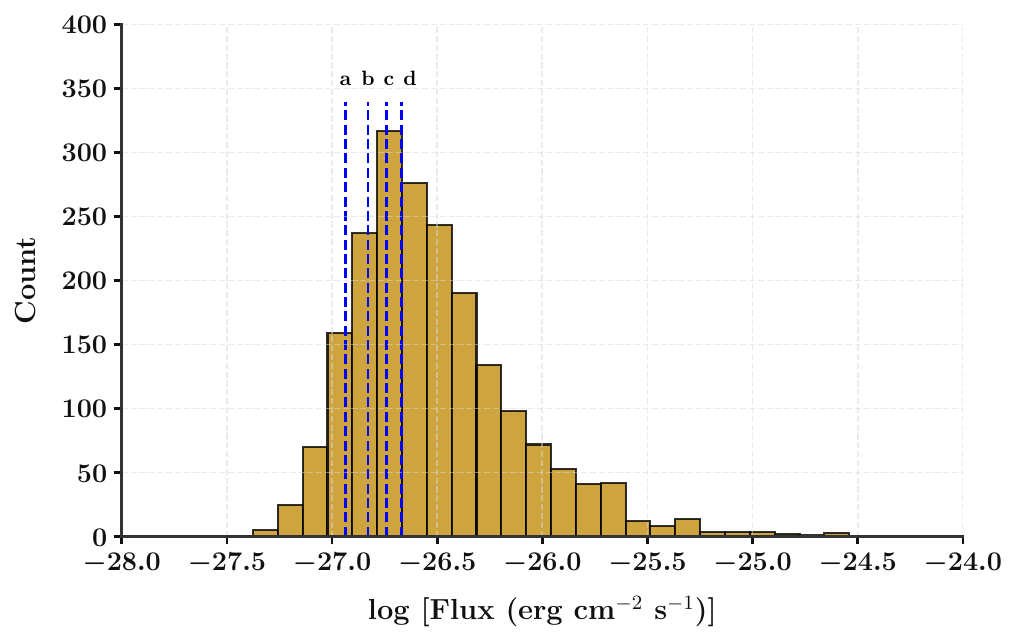}
    \caption{Histogram of the logarithm of radio fluxes of pulsars at 1400 MHz for lfl06 radial distribution generated with psrpoppy. The dashed lines $a$, $b$, $c$, $d$ denote the flux thresholds.
 The thresholds $a$, $b$, $c$, $d$ have been chosen in such
a way that $a$ discards $10\%$, $b$ discards $20\%$, $c$ discards $30\%$, and d discards $40\%$ of the pulsars.}
    \label{fig:histo}
\end{figure}

\begin{figure}[h]
    \centering    \includegraphics[width=0.5\linewidth]{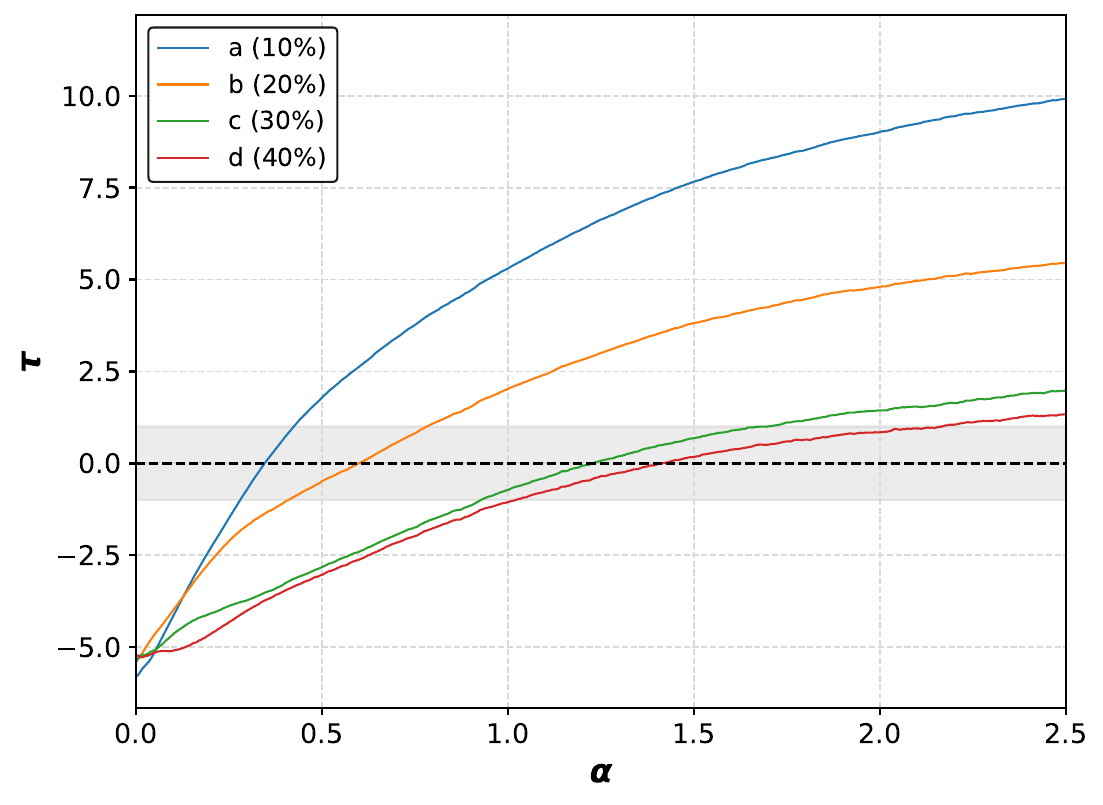}
    \caption{The Efron-Petrosian statistic $\tau$ versus $\alpha$ for different flux thresholds (cf. Fig.~\ref{fig:histo} computed on the synthetic pulsar
dataset generated using PsrPopPy with the lfl06 radial distribution model. The gray shaded region corresponds to  $|\tau|<1$. }
    \label{fig:tauvsalpha}
\end{figure}

We now apply the E-P method to this synthetic population.  The histogram showing the distribution of pulsar fluxes can be found in Fig.~\ref{fig:histo}. Similarly to M23 and ~\cite{Ardavan23}, we consider four different flux thresholds to calculate $\tau$. These thresholds, which we labeled as $a$,$b$,$c$,$d$ were chosen in such a way  that they discard 10\%, 20\%, 30\%, and 40\%, respectively, from the pulsar population.   These four thresholds allow us to test any possible dependence of the E-P test as a function of the flux threshold.
The histogram of the flux distribution for our synthetic sample along with the four  thresholds is shown in Fig.~\ref{fig:histo}. The results for $\tau$ as a function of the distance exponent $\alpha$ for all the four aforementioned thresholds can be found in Fig.~\ref{fig:tauvsalpha}. For a perfect inverse-square law, $\tau$ should be exactly zero for $\alpha=2$ and for other distance exponents should show significant deviations from $\alpha=2$. When applied to real data, because of the uncertainty in distance estimates or other observational systematics, $\tau$ is typically chosen to be between $\pm 1$ in order for the two variables to be uncorrelated~\cite{Dainotti22}. However, since our synthetic pulsar catalog has an inverse-square law dependence for the flux built in and no uncertainties in distance have been incorporated, we should expect $\tau$ to be exactly zero for 
$\alpha=2$. 

\begin{figure}
    \centering
    \includegraphics[width=0.5\linewidth]{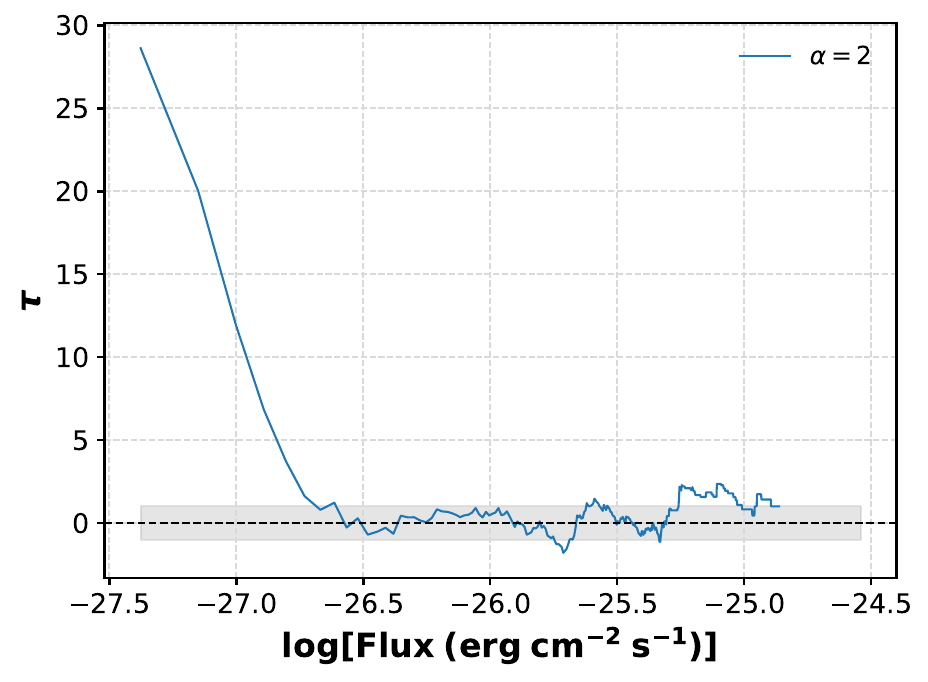}
    \caption{Efron-Petrosian $\tau$ as a function of the  flux threshold plot for a fixed distance exponent of 2.  The gray shaded region corresponds to   $|\tau|<1$. }
    \label{fig:tauvsthreshold}
\end{figure}

However, Fig.~\ref{fig:tauvsalpha} shows that this is clearly not the case.  We find that $\tau$ is greater than 1 for $a$, $b$, and $c$ flux thresholds for $\alpha=2$ and is within 1.0 only for threshold $d$.  For these thresholds, the value of $\tau$ is close to zero for $\alpha$ between $0.5-1.5$. We don't get perfect agreement with $\alpha=2$ for all  the four thresholds. 
We also confirmed that the same is true for  other models of distance distribution in {\tt PsrPopPy}. We also checked the trends of $\tau$ as a function of flux threshold for fixed  $\alpha=2$. This plot can be found in Fig.~\ref{fig:tauvsthreshold}. We find that only for flux thresholds within a narrow range $-26.5 \leq \log \rm{[Flux (ergs~cm^{-2}~s^{-1})]} \leq -25.5$, we obtain values of $\tau$ within $\pm 1$. Even within this flux range, we do not get pristine agreement.
This shows that E-P method fails to recover the correct distance exponent of $\alpha=2$ for our synthetic pulsar population (except over a very small range of flux thresholds), even though the inverse-square law dependence is built in.


\subsection{Diagnosis of the failure of E-P method}
In order to identify the underlying cause as to   why the  E-P method does not work, we first outline the reasons because of which pulsars are culled from the final output in {\tt PsrPopPy}. Candidate pulsars get truncated  from the final sample due to one or more of the following reasons~\cite{PsrPopPy}:
\begin{enumerate}
\item SNR $<$ threshold.
\item Due to  pulse smearing when the effective pulse width is greater than the pulsar period~\cite{handbook}. The effective pulse width is given by the quadrature sum of intrinsic pulse width, sampling time, dispersive smearing across the frequency channel and finally smearing due to free electrons in the interstellar medium. Analytical expressions for each of these can be found in ~\cite{PsrPopPy}. Note, however, that pulsars which get excised because of smearing do not have an assigned  SNR value and therefore have already been removed by the SNR cutoff ~\footnote{Such pulsars have an imaginary value for the SNR in {\tt PsrPopPy.} }
\item Pulsars are located outside the volume of the survey. These pulsars are removed based on the RA,DEC
\end{enumerate}

For the Parkes Multi-beam survey, the value of SNR used in {\tt PsrPopPy} is equal to 9.0. For our sample, out of a total of 2,29,744 pulsars, 1,28,767 were removed because of the SNR cut, while 56,514 and 42,449 were removed due to smearing and  because they were outside the survey area, respectively\footnote{This was obtained from the output of populate.py}. We first do a series of tests to get to the bottom of the problem, which we enumerate below

\subsection{E-P test with zero  SNR cutoff}
We  now modify the output of {\tt PsrPopPy} so that the effective SNR cut-off used  is equal to zero and no pulsars get removed because of the SNR threshold.  For this purpose,  we ran  both {\tt populate} and {\tt dosurvey} with a SNR cut of zero.

\begin{figure}[h]
    \centering
    \includegraphics[width=0.5\linewidth]{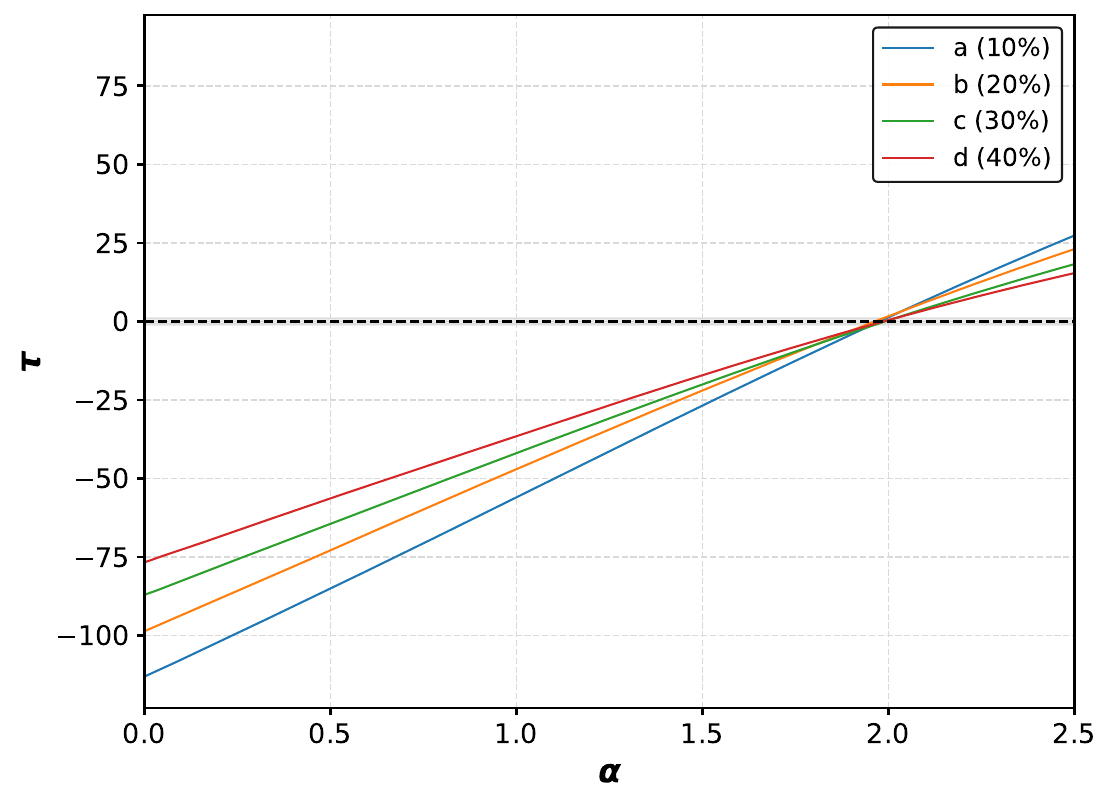}
    \caption{The Efron-Petrosian statistic $\tau$ versus $\alpha$ for different flux thresholds  computed on the synthetic pulsar
dataset   for a SNR cutoff of zero. The gray shaded region corresponds to   $|\tau|<1$. 
We find exact agreement with the inverse-square law for all the four flux thresholds.}
    \label{fig:snrzero}
\end{figure}

All other inputs to {\tt PsrPopPy} were the same as those used for Fig.~\ref{fig:tauvsalpha}. We then run the E-P test on this pulsar dataset in the same way as before. This plot of $\tau$ vs $\alpha$ for different flux thresholds can be found in Fig.~\ref{fig:snrzero}. In this case, $\tau$ is exactly equal to zero for the expected distance exponent of $\alpha=2$ for all the four thresholds, thus implying pristine agreement with the inverse-square law. 
Therefore, we have demonstrated that the E-P method can reproduce the expected inverse-square law scaling of the pulsar flux, when we do not impose a SNR cutoff.

\subsection{Reason for failure of E-P method for non-zero SNR threshold}
\begin{figure}[hbt!]
    \centering
    \includegraphics[width=0.5\linewidth]{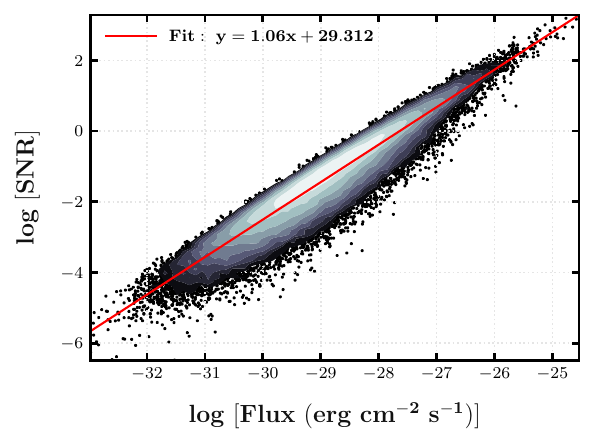}
    \caption{Scatter contour plot showing  SNR  as a function of Flux at 1400 MHz for synthetic pulsars simulated using {\tt PsrPopPy}. The best-fit regression relation is indicated by the solid red line. We find that there is a non-linear relationship between SNR and flux, along with considerable scatter. Note that that the high density regions have been replaced by contours following the prescription in ~\cite{astroML}.}
    \label{fig:snrvsflux}
\end{figure}

Therefore, the analysis in the previous section shows that E-P method works when there is no SNR cutoff imposed. We note that  previously the E-P method has mostly been applied to data which was truncated based on a flux threshold. However, in this case, {\tt PsrPopPy} does not remove faint pulsars based on a flux cut, but instead uses a SNR cut. The SNR of a putative pulsar is given by the radiometer equation~\cite{PsrPopPy,handbook}.  We plot the SNR of our synthetic pulsar population as a function of the flux. This plot can be found in Fig.~\ref{fig:snrvsflux}. We find that there is an inherent non-linear relationship between the SNR and flux. We can fit a regression relation (by binning the flux and using the median SNR in each bin) using 
\begin{equation}
\log (\text{SNR}) =  1.06\log(\text{Flux})+ 29.312,
\label{eq:SNRvsflux}
\end{equation}
where Flux is expressed in ergs $\rm{cm^{-2}~s^{-1}}$. 
Furthermore, in addition to the non-linear relation there is also considerable scatter around this best-fit.  Consequently, these two factors cause the E–P method to fail, as the SNR threshold is raised. We are not aware of any previous application of E-P  on a dataset,  which  was truncated using  a variable which  shows a non-linear relationship with flux  along with  considerable scatter.

\begin{figure}[h]
    \centering
    \includegraphics[width=0.5\linewidth]{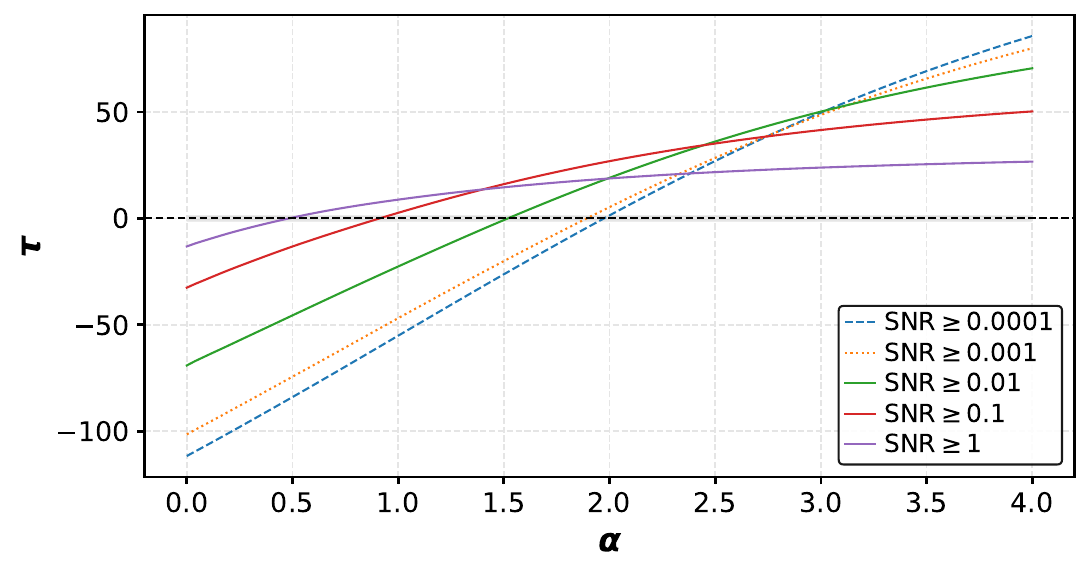}
    \caption{Efron-Petrosian $\tau$ vs distance exponent ($\alpha$) for SNR  cut-off values of 0.0001,0.001, 0.01, 0.1 and 1 values. The previously defined flux threshold of $a$ is applied here. We can see that increasing the SNR cut-off leads to larger deviations of $\tau$ from its expected value of zero at $\alpha=2$.
    Conversely, the value of $\tau$ approaches zero for increasingly small distance exponents as the SNR cut-off is raised.}
    \label{fig:vssnr}
\end{figure}

To further elucidate this, we perform the E-P test for different pulsar datasets selected according to different SNR cut-offs. For each dataset, we plot $\tau$ as a function of $\alpha$ for one fixed  threshold $a$, which is chosen
according to the criterion in Fig.~\ref{fig:histo}. This plot can be found in 
Fig.~\ref{fig:vssnr}. We find that for very small values of SNR, $\tau$ is close to 0 for $\alpha=2$, corresponding to inverse-square law. However, as we increase the SNR cut-off, we observe larger  deviations of $\tau$ from its expected value of zero for $\alpha=2$. Conversely, the value of $\tau$ approaches zero for increasingly small distance exponents as the SNR cut-off is raised.

\section{E-P Test using  a flux-based cutoff}
\label{sec:fluxcut}
\begin{figure}[h]
     \centering
     \begin{subfigure}[b]{0.3\textwidth}
         \centering
         \includegraphics[width=\textwidth]{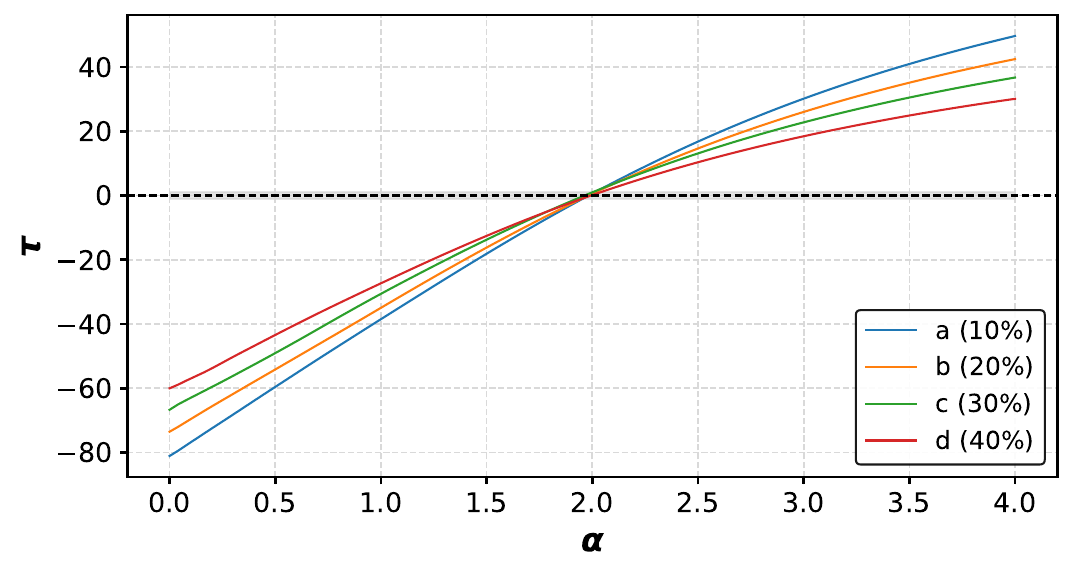}
         \caption{$\geq$ Flux (SNR=0.01)}
         \label{fig:flux_cut_0.01}
     \end{subfigure}
     \hfill
     \begin{subfigure}[b]{0.3\textwidth}
         \centering
         \includegraphics[width=\textwidth]{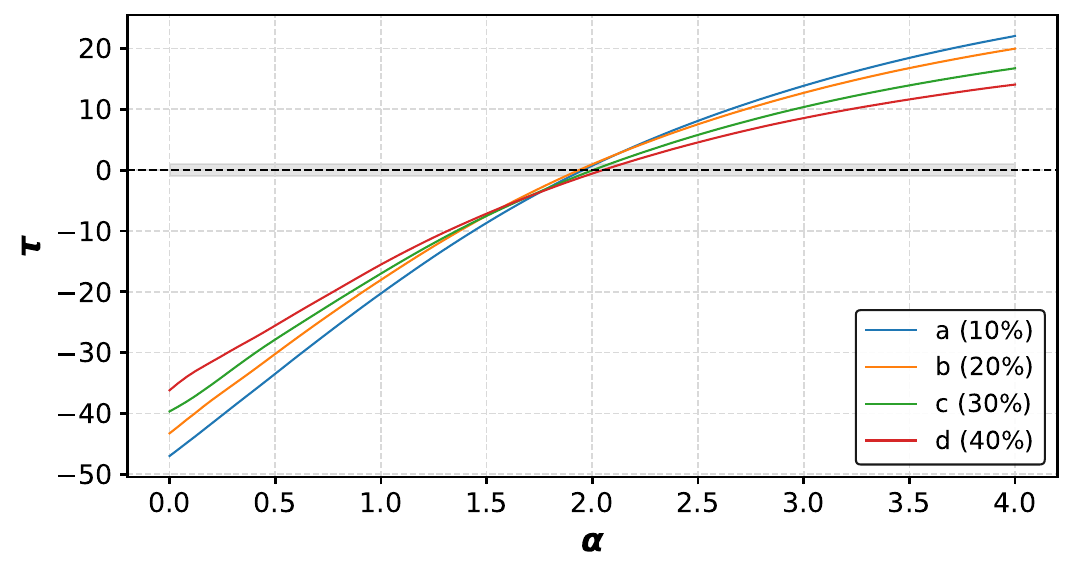}
         \caption{$\geq$ Flux (SNR=0.1)}
         \label{fig:flux_cut_0.1}
     \end{subfigure}
     \hfill
     \begin{subfigure}[b]{0.3\textwidth}
         \centering
         \includegraphics[width=\textwidth]{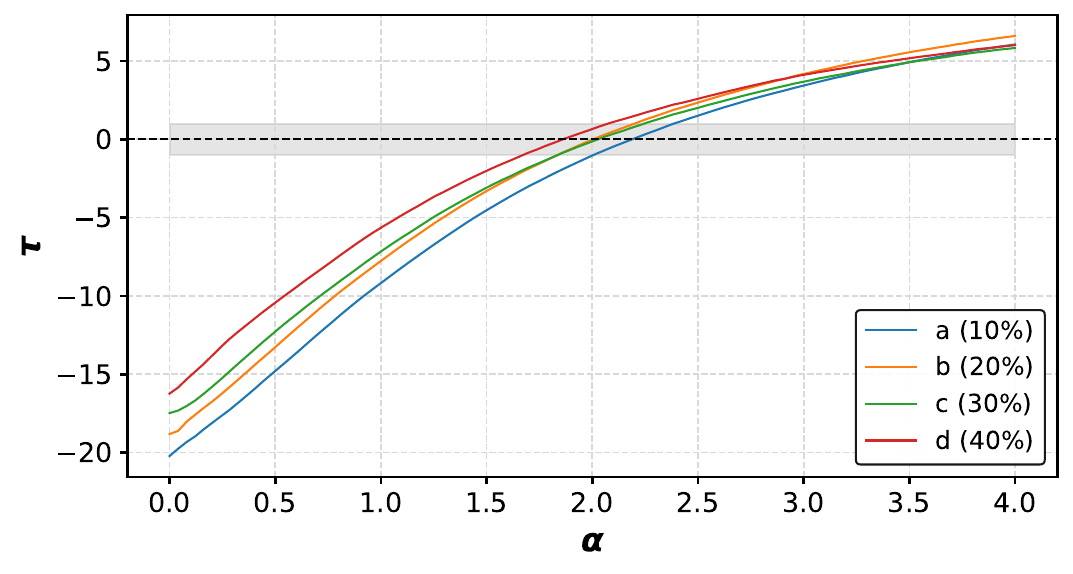}
         \caption{$\geq$ Flux (SNR=1)}
         \label{fig:flux_cut_1}
     \end{subfigure}
        \caption{E-P   $\tau$ versus $\alpha$ for a synthetic pulsar catalog which is chosen on the basis of a flux threshold, instead of a SNR-based cutoff. The flux thresholds correspond to SNR of 0.01,0.1,1 respectively on the basis of regression relation in Eq.~\ref{eq:SNRvsflux}. The gray shaded region corresponds to  $|\tau|<1$. }
        \label{fig:flux_cut}
\end{figure}

We now check whether we can recover the correct distance scaling by truncating the dataset based on the flux. 
Therefore, based on what we have deduced before, we now create a new synthetic sample, where we excise pulsars from the sample based on a flux cut, instead of SNR. We then rerun E-P tests using the new synthetic catalog thus constructed. These plots for three different flux thresholds corresponding to SNRs of 0.01, 0.1, and 1.0 when using the regression  relation in Eq.~\ref{eq:SNRvsflux} can be found in Fig.~\ref{fig:flux_cut}. We find that for  flux values corresponding to SNRs of 0.01 and 0.1, we obtain pristine agreement with the inverse-square law scaling for the pulsar flux. For flux threshold corresponding to a SNR of 1.0,  we get $|\tau|<1$, implying that  the agreement with the inverse-square law is within $\pm 1 \sigma$.

\begin{figure}[h]
     \centering
     \begin{subfigure}[b]{0.3\textwidth}
         \centering
         \includegraphics[width=\textwidth]{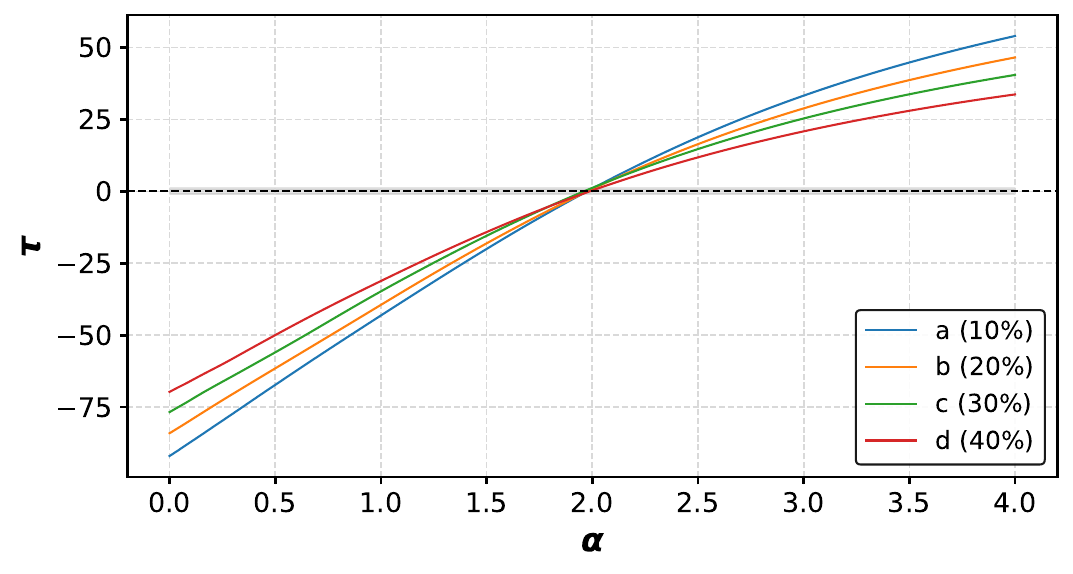}
         \caption{$\geq$ Flux (SNR=0.01)}
         \label{fig:flux_cut_0.01smear}
     \end{subfigure}
     \hfill
     \begin{subfigure}[b]{0.3\textwidth}
         \centering
         \includegraphics[width=\textwidth]{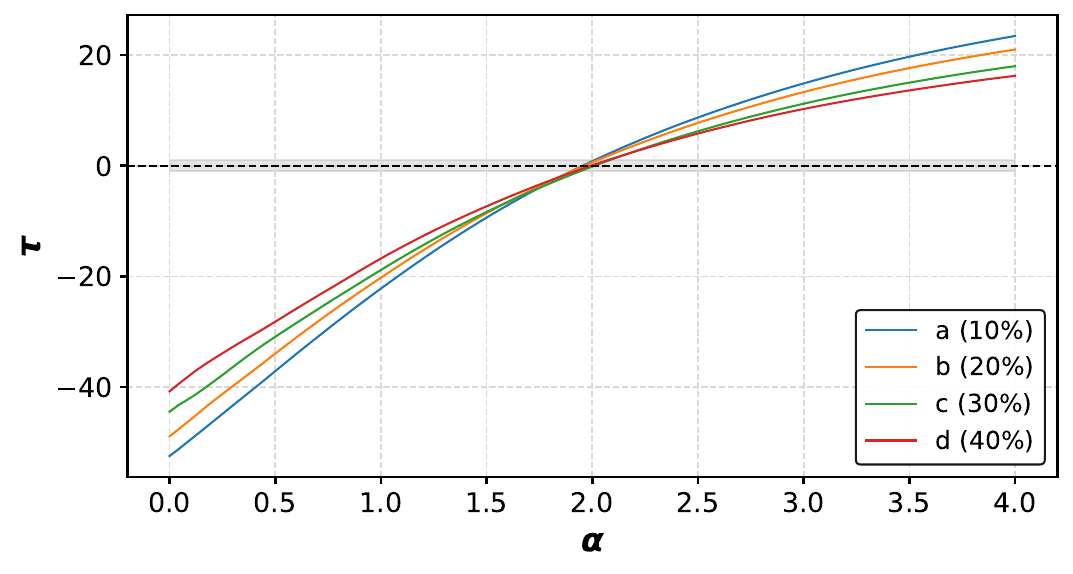}
         \caption{$\geq$ Flux (SNR=0.1)}
         \label{fig:flux_cut_0.1smear}
     \end{subfigure}
     \hfill
     \begin{subfigure}[b]{0.3\textwidth}
         \centering
         \includegraphics[width=\textwidth]{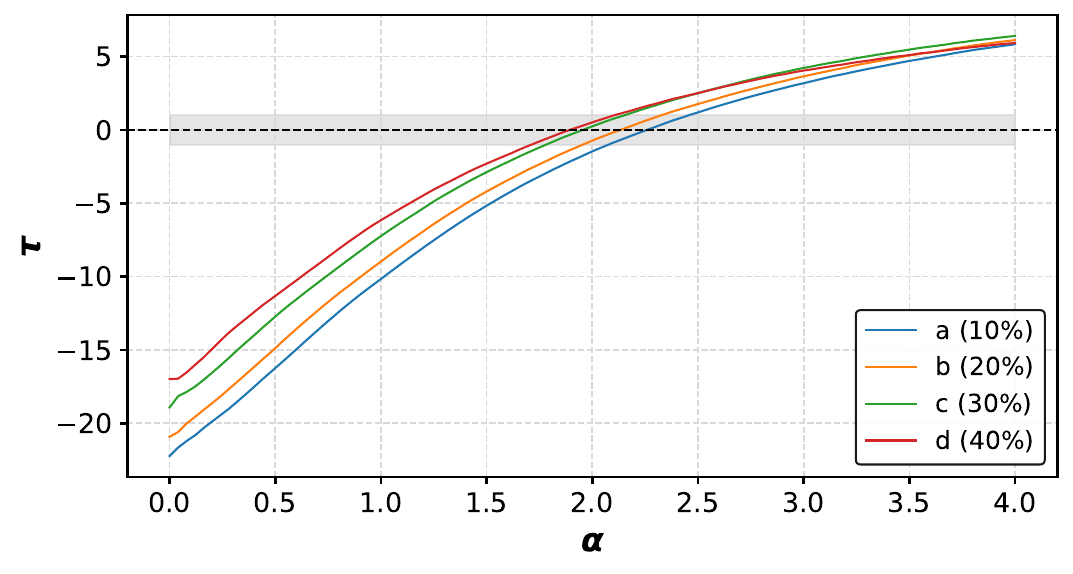}
         \caption{$\geq$ Flux (SNR=1)}
         \label{fig:flux_cut_1smear}
     \end{subfigure}
        \caption{ E-P   $\tau$ versus $\alpha$ for a synthetic pulsar catalog including the smeared population which was  chosen on the basis of a flux threshold, instead of a SNR-based cutoff. The flux thresholds correspond to SNR of 0.01,0.1,1 respectively on the basis of regression relation in Eq.~\ref{eq:SNRvsflux}. The gray shaded region corresponds to   $|\tau|<1$. 
        }
        \label{fig:flux_cutwithsmeared}
\end{figure}

In order to check the effect of the smeared population in the results, we also tried  to include the smeared population  and redid the above test. We note that SNR is not defined for pulsars which get culled due to smearing.   However, the output of {\tt PsrPopPy} includes   a measured flux for such pulsars which otherwise get removed due to smearing. Therefore, we  manually append the smeared pulsars to the existing flux-selected catalog based on their flux values. These plots for the same flux thresholds as above  can be found in Fig.~\ref{fig:flux_cutwithsmeared}. We find that the results are almost the same as those without including the smeared pulsars. Therefore, the pulsars which are culled due to smearing do not affect the E-P test.


Therefore, we find that the reason for the failure of the E-P method to recover the inverse-square law scaling is due to the fact that the synthetic pulsar dataset is truncated based on a SNR cutoff, and SNR has an inherent non-linear relationship with  flux along with  considerable scatter.  Although we are not aware of a similar use case in literature, it has been pointed out, that incorrect conclusions have been obtained regarding the Gamma-ray burst (GRB) luminosity function  using the E-P method based upon a single flux threshold, even though the detection threshold of these detectors is more complex than a single cutoff~\cite{Bryant}.

\section{Conclusions}
\label{sec:conclusions}
In a recent work (M23), we had applied the E-P method to radio pulsar flux at 1400 MHz on the data from Parkes Multi-beam survey in order to ascertain if we could find a deviation from inverse-square law in the radio pulsar flux as claimed in one of the theoretical models for pulsar emission~\cite{Ardavan08}, which had hitherto been claimed for X-ray and gamma-ray fluxes of pulsars and magnetars. No firm conclusion could be drawn from the analysis in M23.

In this work, we follow up on that analysis and try to demonstrate if we can recover the inverse-square law scaling of the flux for a synthetic radio pulsar population. We use  a controlled setup in which   the inverse-square law scaling of the pulsar flux is built in by construction, and the pulsar distances are known exactly, thereby eliminating any distance uncertainties.

As a first step, we generate a simulated dataset  having a lognormal distribution in luminosity according to  the specifications of the Parkes Multi-beam survey using the {\tt PsrPopPy} software. We find that the E-P method cannot recover the inverse-square law scaling of the pulsar flux for most thresholds used. Only for a small subset of thresholds (for which over 40\% of the sample is culled), do we get consistency with the inverse-square law to within $1\sigma$.

We then performed   a series of tests to get to the bottom of the problem. We first showed that we get pristine agreement with the inverse-square law, when we do not  impose any cut based on SNR.
We then showed that the disagreement with inverse-square law becomes more pronounced, as we raise the SNR cutoff. We then showed that there is a non-linear relation between the SNR and the flux along with a considerable scatter. We then constructed a new synthetic dataset which was truncated according to a fixed flux threshold instead of SNR. We then demonstrated that only with such a flux-based cut, we can recover the true distance exponent. This agreement is exact for lower values of flux cutoff.  However, at higher values of the flux cutoff, the agreement with inverse-square law is only within $\pm 1\sigma$.  In Appendix A we have shown that our conclusions do not change for  a power-law distribution for the pulsar luminosity. In Appendix B, we have also shown that the E-P method also cannot recover the correct scalings for a synthetic pulsar population having inverse linear and inverse cubic scalings with distance. 

Therefore, one cannot use the E-P method for testing the correct inverse-square law scaling of radio pulsar flux with distance. Furthermore, we have also shown that the E-P  method breaks down and  cannot be used to determine  scaling relations when the detection threshold does not scale linearly with the observed flux, and in addition has scatter around this non-linear relation. In future works we shall perform  similar tests with synthetic gamma-ray fluxes of pulsars.

In the spirit of open science we have made our analysis codes and synthetic dataset publicly available, which can be found at \url{https://github.com/sanjith123456/EFRON-PETROSIAN}.

\section*{Acknowledgements}
We are grateful to Pragna Mamidipaka for making the analysis codes in M23 publicly  available. We also thank the anonymous referee for several constructive comments and feedback on our manuscript. This work is also dedicated to the memory of Prof. Houshang Ardavan, whose recent papers on testing scaling of pulsar flux with distance  motivated us to look into this issue in detail.

\bibliography{main}

\appendix

\section{Tests with a power-law distribution of luminosity}
We now rerun the E-P method on a synthetic sample  of pulsars having a power-law distribution of luminosity instead of the lognormal distribution as done in the main manuscript. For this purpose, we  used a power-law index of -0.7 for the luminosity exponent based on the model  in ~\cite{Lorimer06}. To generate the power-law, we ran {\tt populate} with the additional command line options {\tt --ldist pow -l 0.00924 559.53 -0.7}, where the first and second arguments refer to the minimum and maximum luminosity in units of mJy kpc$^2$, and the third argument refers to the power-law index for the luminosity.

\begin{figure}[h]
    \centering
    \includegraphics[width=0.5\linewidth]{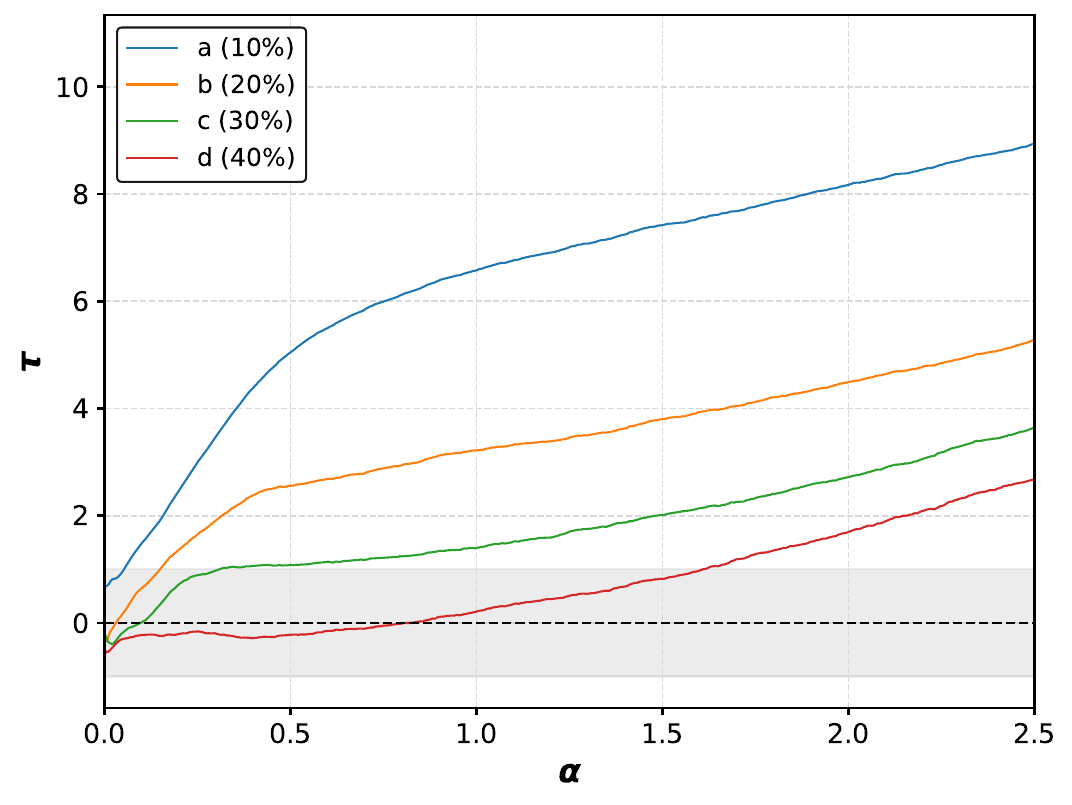}
    \caption{E-P $\tau$ as a function of  $\alpha$ for different flux thresholds (cf. Fig.~\ref{fig:histo}) computed on the synthetic pulsar dataset generated using a  power-law distribution for the luminosity. The gray shaded region  corresponds to  $|\tau|<1$.}
    \label{fig:default_pow}
\end{figure}

\begin{figure}[h]
    \centering
    \includegraphics[width=0.5\linewidth]{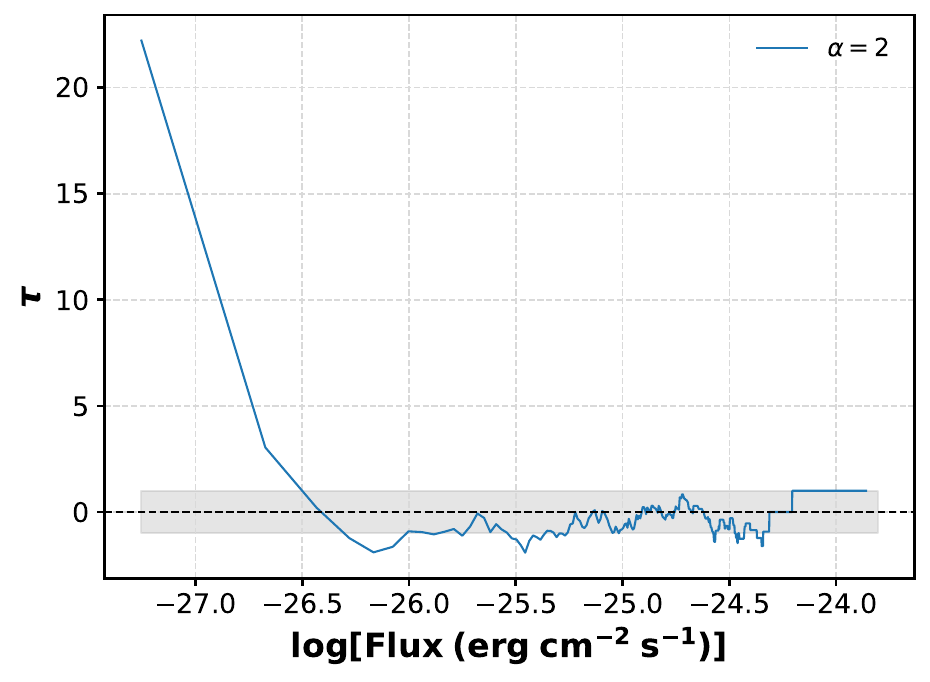}
    \caption{E-P $\tau$ as a function of the  flux threshold  for a fixed distance exponent of 2 for luminosity law distribution of luminosity.  The gray shaded region  corresponds to  $|\tau|<1$.
}
    \label{fig:tauvsthresh_pow}
\end{figure}

We then run the E-P test in the same way as in the main manuscript.  Our results for $\tau$ as a function of $\alpha$ can be found in Fig.~\ref{fig:default_pow} for the same four flux thresholds as before. We find that for all four thresholds, the value of $|\tau|$ is $>1$ for $\alpha=2$.  Therefore, this shows that even with a power-law distribution of luminosities, we cannot recover the correct inverse-square law scaling with distance for the radio pulsar flux. When we plot $\tau$ as a function of flux threshold for  $\alpha=2$ (cf. Fig.~\ref{fig:tauvsthresh_pow}), we find values of $\tau$ within $\pm 1.0$ only for 
 $\log \rm{[Flux (ergs~cm^{-2}~s^{-1})]} \geq -26.0$, and similar to before, the agreement is not pristine.

 Therefore, the results of the E-P test for a synthetic sample with power-law distribution of luminosity are comparable to those for a lognormal distribution.

\section{Tests with synthetic fluxes having  different scalings with distance}

We now generate synthetic pulsar fluxes with $F \propto 1/d$ and
$F \propto 1/d^3$ with the remaining settings same as in the main manuscript. We first  run {\tt populate.py} and {\tt dosurvey.py} in the same way in Sect.~\ref{sec:simulations}.
We then  performed E-P analysis on these outputs following the same prescription as Sect.~\ref{sec:analysis}.
We then test if the E-P method can recover the correct distance exponents of $n=1$ and $n=3$ for both the cases.
To generate synthetic fluxes for these scales, we modified {\tt PsrPopPy} so that the output flux scales with luminosity according to
$F=L/D^{\alpha} l^{2-\alpha}$ (cf. Eq.~\ref{eq:L}), where the  constant value of $l$ was used so that the new flux has the same dimensions as that for an inverse-square law. We use the median distance of our synthetic pulsar population, given by $l =4.6$ kpc.

\begin{figure}
    \centering
    \includegraphics[width=0.5\linewidth]{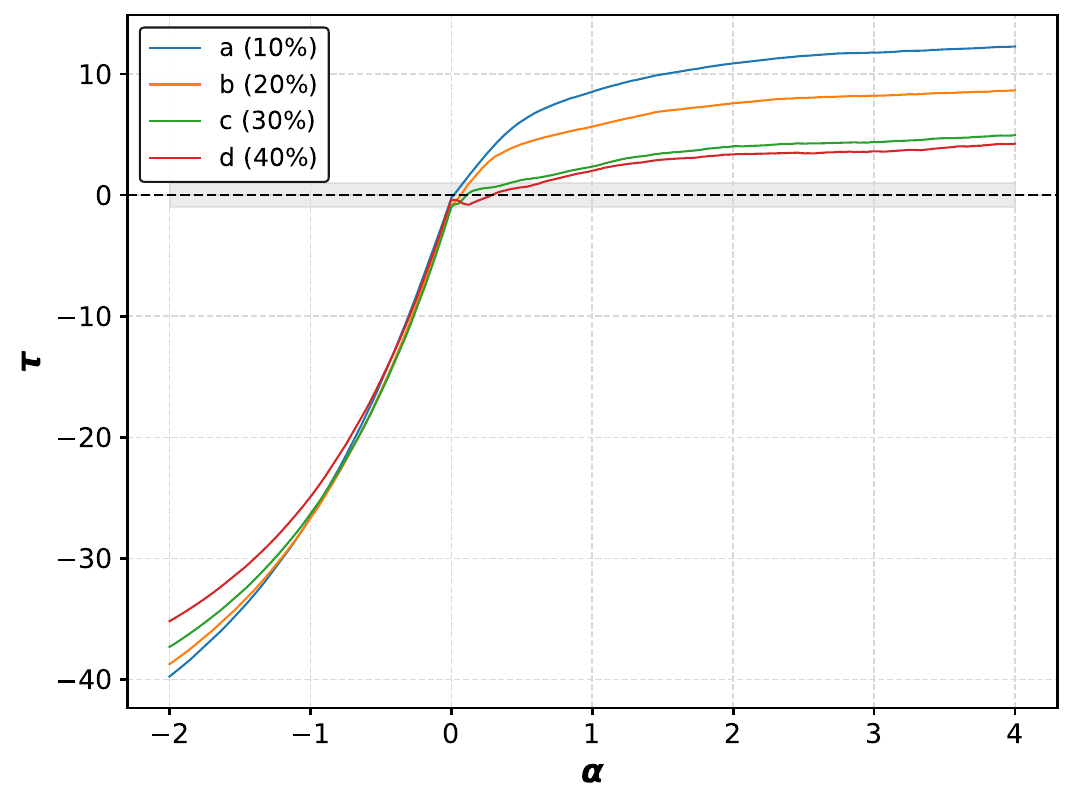}
    \caption{
    E-P $\tau$ as a function of  $\alpha$ for different flux thresholds (cf. Fig.~\ref{fig:histo}) for a synthetic pulsar population where $F \propto 1/ D$.  The gray shaded region corresponds to   $|\tau|<1$.  We find that E-P cannot recover the true distance exponent $\alpha=1$ for all values of the flux threshold. }
    \label{fig:Tauvsalpha_alpha_1_l}
\end{figure}

\begin{figure}
    \centering
    \includegraphics[width=0.5\linewidth]{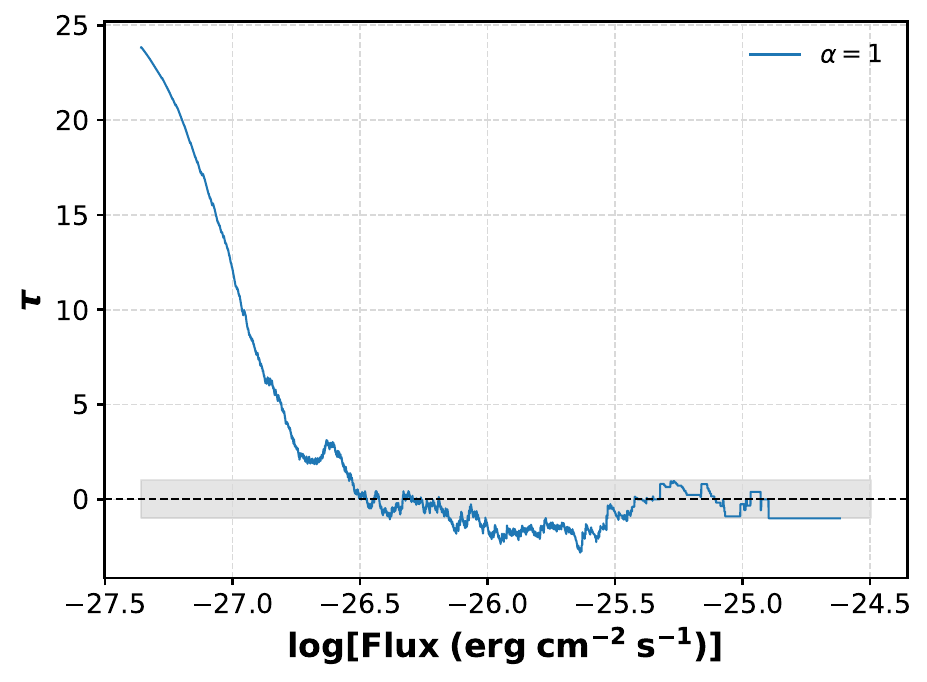}
    \caption{E-P $\tau$ as a function of the  flux threshold  for a fixed distance exponent of 1 for synthetic pulsar population, where flux scales inversely with distance ($F \propto 1/D$). The gray shaded region shows the $\pm 1 $ range for $\tau$.}
    \label{fig:Tauvsthresh_alpha_1_l}
\end{figure}

These plots of $\tau$ as a function of $\alpha$ and flux threshold (assuming $\alpha=1$) for inverse-distance  scaling of flux 
can be found in Fig.~\ref{fig:Tauvsalpha_alpha_1_l} and ~\ref{fig:Tauvsthresh_alpha_1_l}, respectively. We find that the E-P cannot recover the true distance exponent ($\alpha=1$) for this synthetic population where the flux scales inversely with distance. The trends for these plots are similar to those for the inverse-square law variation with distance.


\begin{figure}
    \centering
    \includegraphics[width=0.5\linewidth]{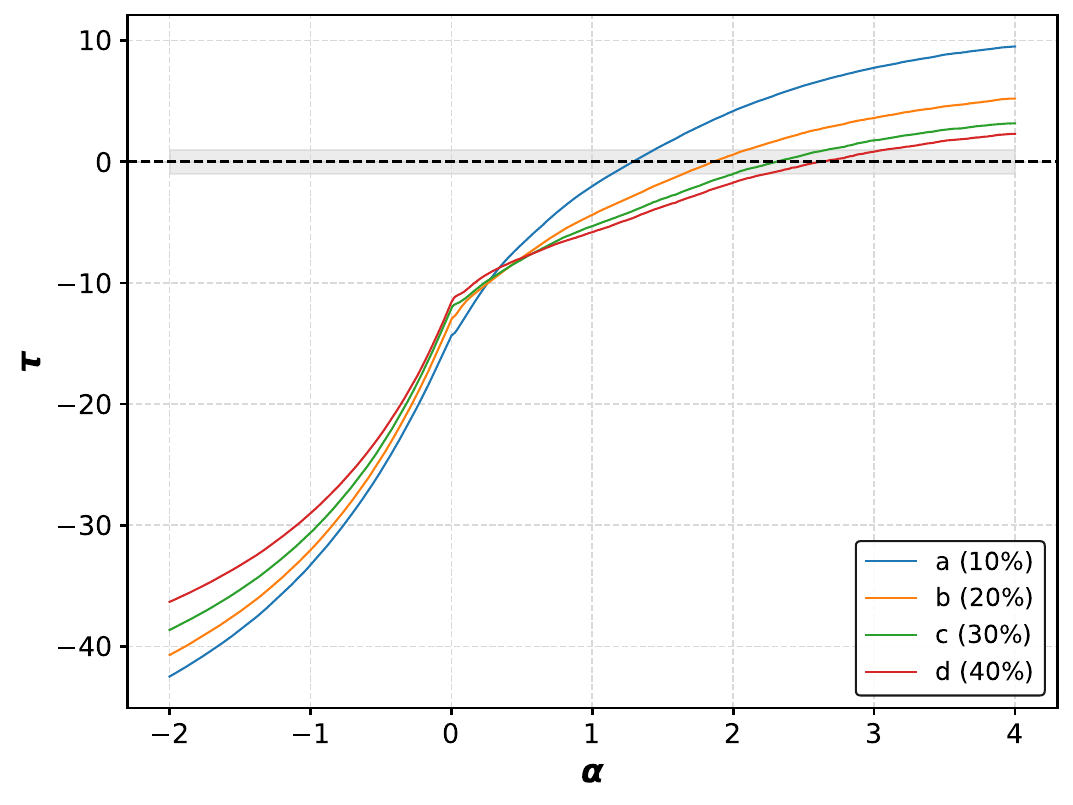}
    \caption{
    E-P $\tau$ as a function of  $\alpha$ for different flux thresholds (cf. Fig.~\ref{fig:histo}) for a synthetic pulsar population where $F \propto 1/ D^3$.  The gray shaded region corresponds to  $|\tau|<1$.
    We find that E-P cannot recover the true distance exponent $\alpha=3$ for most values of the flux threshold. }
    \label{fig:tauvsalpha_3_l}
\end{figure}

\begin{figure}
    \centering
    \includegraphics[width=0.5\linewidth]{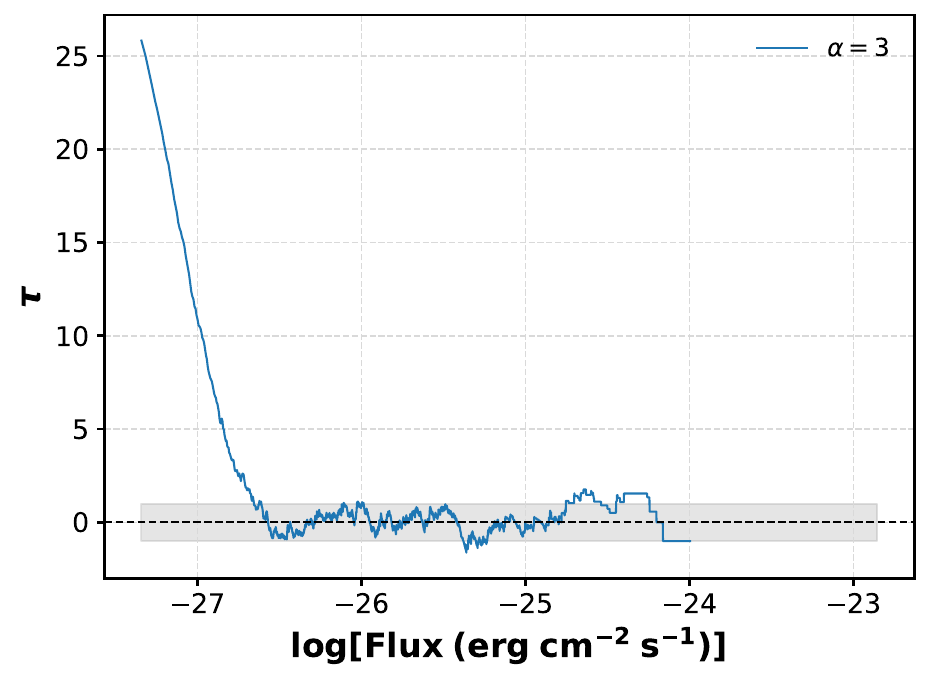}
    \caption{E-P $\tau$ as a function of the  flux threshold  for a fixed distance exponent of 3 for synthetic pulsar population, where flux scales inversely with distance ($F \propto 1/D^3$). The gray shaded region corresponds to  $|\tau|<1$.}
    \label{fig:tauvsthresh_3_l}
\end{figure}

The corresponding plots for synthetic pulsar fluxes having $F \propto 1 /D^3$ can be found in Fig.~\ref{fig:tauvsalpha_3_l} and Fig.~\ref{fig:tauvsthresh_3_l}. Once again, we find that we cannot recover the correct distance exponent ($\alpha=3.0$) for most values of the flux threshold. Similarly the $\tau$ versus  Flux threshold plot for the correct distance exponent also shows similar trends as for the inverse square law scaling, making it impossible to discern the correct distance exponent.

Therefore, we have shown that E-P method cannot recover the correct distant exponents when synthetic fluxes are generated using scalings that differ from  the inverse-square law.


\end{document}